# Decoding Working Memory Load from EEG with LSTM Networks

Samuel Goldstein[1,2], Zhenhong Hu[2], Mingzhou Ding[2]*

[1] Department of Electrical & Computer Engineering, University of Florida, Gainesville, Florida, 32611

[2] J. Crayton Pruitt Family Department of Biomedical Engineering, University of Florida, Gainesville, Florida, 32611

**\* Corresponding Author:** Mingzhou Ding

Phone: +1-352-273-9332

Fax:   +1-352-273-9221

Email: MDing@bme.ufl.edu

Number of Pages: 30

Number of Figures: 6

Number of Tables: 0

Number of Words for Abstract: 254

# Abstract

Working memory (WM) is a mechanism that temporarily stores and manipulates information in service of behavioral goals and is a highly dynamic process. Previous studies have considered decoding WM load using EEG but have not investigated the contribution of sequential information contained in the temporal patterns of the EEG data that can differentiate different WM loads. In our study, we develop a novel method of investigating the role of sequential information in the manipulation and storage of verbal information at various time scales and localize topographically the sources of the sequential information based decodability. High density EEG (128-channel) were recorded from twenty subjects performing a Sternberg verbal WM task with varying memory loads. Long Short-Term Memory Recurrent Neural Networks (LSTM-RNN) were trained to decode memory load during encoding, retention, activity-silent, and retrieval periods. Decoding accuracy was compared between ordered data and a temporally shuffled version that retains pattern based information of the data but not temporal relation to assess the contribution of sequential information to decoding memory load. The results show that (1) decoding accuracy increases with increase in the length of the EEG time series given to the LSTM for both ordered and temporally shuffled cases, with the increase being faster for ordered than temporally shuffled time series, and (2) according to the decoding weight maps, the frontal, temporal and some parietal areas are an important source of sequential information based decodability. This study, to our knowledge, is the first study applying a LSTM-RNN approach to investigate temporal dynamics in human EEG data in encoding WM load information.

# Introduction

Working memory (WM) is a system for the temporary storage and manipulation of information in service of behavioral goals and is a highly dynamic process involving multiple regions of the brain. WM is comprised of the following components: central executive, episodic buffer, phonological loop (verbal WM), and visuospatial sketchpad (visual WM) (Baddeley, 2003). There are multiple processing stages in a typical WM task: encoding during which neural representations of external stimuli are created, retention during which stimulus information is maintained online, activity-silent period during which stimulus information is maintained online without elevated neuronal firing (Stokes, 2015), and retrieval during which stimulus information is retrieved to generate a behavioral response.

The amount of information to be remembered during a WM task is called the memory load. Observing brain activities under different levels of WM load is a commonly applied technique for uncovering the neural mechanisms of WM. Exploiting differences in patterns of neural activity, a machine learning algorithm can be trained on neural imaging data to classify WM load at variable time points of interest. The majority of previous decoding studies are only trained on pattern based information at a single time point (Christophel et al., 2012; LaRocque et al., 2012; Majerus et al., 2016; Mallett and Lewis-Peacock, 2018). This is limiting because WM, especially verbal WM, sequentially stores and manipulates information. The sequential relationship should thus contain clues of the WM load. A few previous decoding studies have applied deep recurrent neural networks with

Long Short-Term Memory (LSTM) to classify WM load. Although these studies strive to preserve spatial, temporal, and spectral information in the EEG data to decode WM load (Bashivan et al., 2015; Kuanar et al., 2018), their main purpose is to optimize the classifier, and not to focus on the neuroscientific interpretation of the results.

We examined whether temporally ordered neural data contributes to the neural representations of WM by applying a LSTM recurrent neural network (LSTM-RNN) to learn robust temporal patterns in sequential frames and predict the levels of WM load from EEG data recorded in humans performing a verbal WM task. We developed a novel approach where decoding accuracy was evaluated for varying time series lengths for an ordered and temporally shuffled scenario with the spatial pattern of the input data preserved to evaluate the contribution of sequential temporal information to decoding memory load across varying temporal scales. To reveal topographically how sequential and pattern based information vary across WM stages, the trained LSTM weights were used to localize brain regions where sequential and pattern based information contribute to decoding accuracy. The analysis is completed for encoding, retention, activity-silent, and retrieval stages of the task to assess how temporal and spatial information varies across WM processes.

# Methods

## Participants

The University of Florida Institutional Review Board approved the protocol for the experiment. Twenty healthy subjects (11 females, 9 males, mean age: 23.55±3.35 years), with normal or corrected-to-normal vision and free from psychiatric or neurological disorders participated in the experiment. Prior to participation, all subjects signed written informed consent.

## Experimental Paradigm

The 20 subjects performed the Sternberg working memory task shown in Figure 1A. For each trial, a fixation cross presented at the center of the screen for 1 s. Then, a memory set was presented for 2 s (encoding). The memory set had an equal probability of containing 2, 4, or 6 upper case consonants arranged in six possible locations in a circle around the fixation point. In order to make the sensory input for the three memory load conditions comparable, the letter X acts as a placeholder for memory sets with 2 or 4 consonants. Memory set varied randomly from trial to trial. A 3-second retention period followed the offset of the memory set. At the end of the retention period, a lower case consonant was shown on the center of the screen for 1 second the subject pressed a button to indicate whether the consonant had occurred in the memory set (retrieval). There was an equal probability that the consonant was or was not part of the memory set. Subjects were instructed to respond as accurately as possible. For each session, there were 3 blocks with 72 trials in each block. The three memory loads were equally likely to occur. Breaks were given between blocks. The experiment lasted 30 minutes. To minimize the effect of learning, a practice session was given prior to EEG recording.

Each subject performed 3 sessions of the experiment as part of a tACS stimulation study (Hu et al., 2019). In addition, they also completed an OSPAN task to measure working memory capacity (WMC).

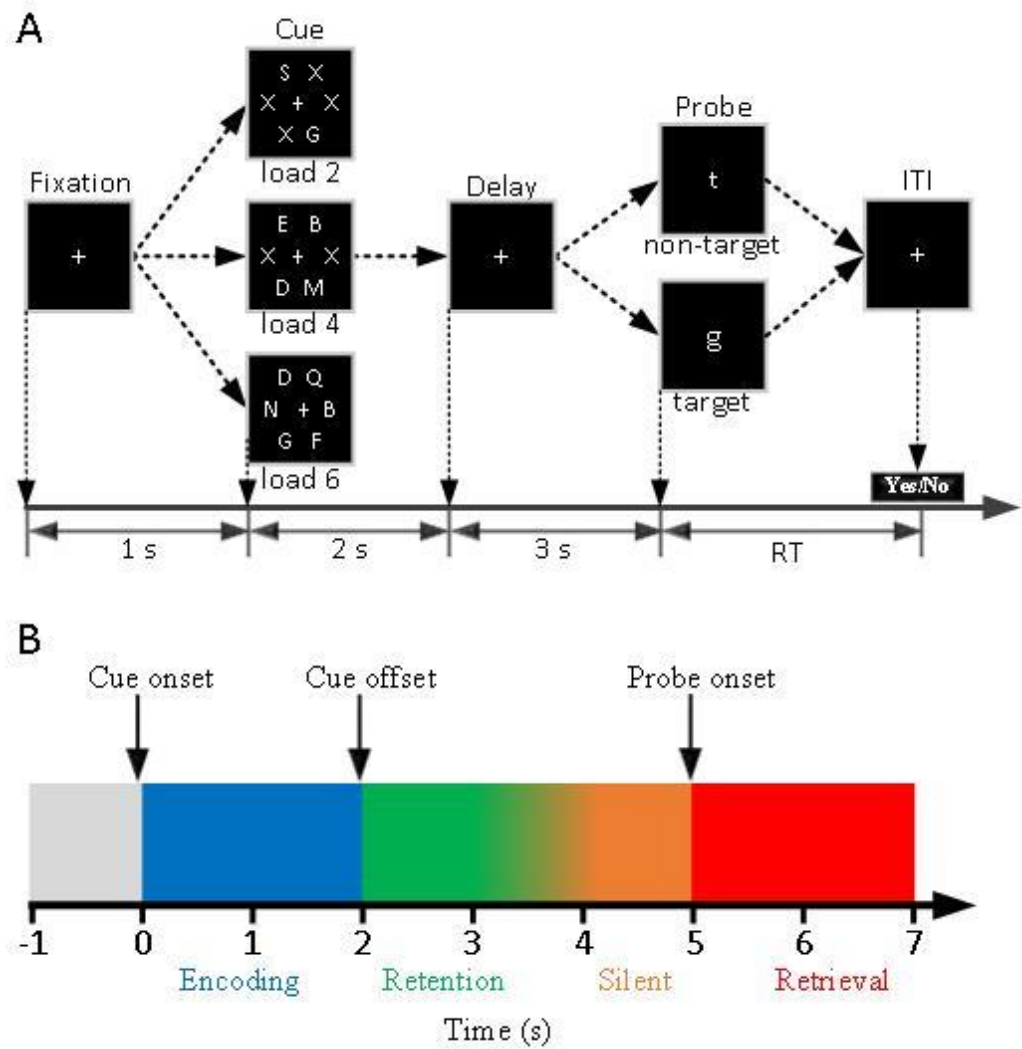

Figure 1. The experiment. (A) Paradigm for the verbal working memory task. (B) Time period of interest for the working memory stages.

## Data acquisition and preprocessing

High-density EEG data was recorded while the subject was performing the Sternberg task. The recording took place in an electrically and acoustically shielded room with a

128-channel BioSemi Active Two System at a 1 KHz sampling rate. Data preprocessing was performed off-line with EEGLAB. The data was high-pass filtered at 0.1 Hz and low-pass filtered at 30 Hz with zero-phase FIR filters. After filtering, the data was downsampled to 256 Hz, re-referenced against the average reference, and epoched from -1 s to 7 s with cue onset occurring at 0 s. Trials with incorrect large movement-related artifacts or incorrect responses were excluded from further analysis. For the remaining trials, independent components analysis (ICA) (Delorme and Makeig, 2004) was applied to remove artifacts due to eye movements, eye blinks, and other noise not related to brain activity. To eliminate the negative effects of volume conduction and common reference, the artifact-corrected scalp voltage data were converted to reference-free current source density (CSD) by calculating 2D surface Laplacian algorithm (Kayser and Tenke, 2006). All further analyses were performed on the CSD data. For all decoding analysis, the CSD time series was temporal smoothed with a length 5 Gaussian kernel. Additionally, the mean and variance of the CSD was standardized across all electrodes.

## SVM Decoding

Support Vector Machines (SVM) has been used to decode working memory conditions in previous studies (Bae and Luck, 2018; Christophel et al., 2012; Esterman et al., 2009). Our first step in this study was to apply SVM to the EEG data to identify time periods of interest (TOI) for further analysis with LSTM-RNN. Load 2 vs Load 6 was decoded at each time point to yield a function of time called the decoding accuracy function. SVM implementation is based on libsvm (Chang and Lin, 2011). A linear kernel was used with the misclassification penalty parameter set to 1. 5-fold cross-validation was implemented. Each subject had an average of 65 ± 7 trials per memory load condition due to artifact removal, across 2 conditions, Load 2 and Load 6, for each of the 3 sessions. Therefore, there is an average of 65*2*3=390 samples used for decoding each time point.

## LSTM Decoding

The LSTM used is a many-to-one recurrent neural network shown in Figure 2A. This type of LSTM is most appropriate as the load condition is static across time for a given trial. The architecture of the LSTM is shown in Figure 2B. Fifty LSTM cells were used because additional cells had marginally diminishing increases in testing decoding accuracy and would be too computationally expensive. Figure 2C shows the architecture inside the LSTM. The LSTM consists of three gates with a sigmoid nonlinearity, "input", "output" and "forget" as well as an "activation" gate with a hyperbolic tangent nonlinearity. These four gates have weight parameters associated with them that learn nonlinear functions in the data. Figure 2D shows mathematically how the gates interact with each other. Each LSTM cell contains an internal state which recurrently updates based on new information entering through the input gate and previous information entering through the forget gate. The activation gate is the primary activation function of the cell with a full dynamic range between -1 and 1 which is operated on by these other gates. The input gate protects the internal state from perturbations from irrelevant inputs. The output gate protects other LSTM cells from perturbations from irrelevant memory information contained in the internal state. The forget gate learns relevant information in the time series to keep as the internal state updates. All four gates have weights associated with recurrent connections among LSTM cell and feedforward connections from the input layer to the LSTM cells which work together to learn pattern-based and sequential information from the EEG signal related to working memory load (Gers et al., 2000; Hochreiter and Schmidhuber,

1997). The LSTM layer was fed into a softmax output layer with a one hot encoding for Load 2 and Load 6 labels. Testing accuracy was evaluated with 5-fold cross validation. For each fold, the LSTM was trained with a batch size of 32 samples and a maximum of 100 epochs. If testing accuracy started to decrease with increased epochs, training would automatically stop to avoid overfitting. The weights are randomly initialized and the biases are initialized to zero. Cross-entropy is used as the loss function for the network. Adam is used to optimize the weights of the LSTM network (Kingma and Ba, 2014). A stateless LSTM was implemented where the internal state of the LSTM cells reset for new batches.

Based on the SVM decoding results shown in Figure 3, four TOIs are assigned to do further decoding analysis with LSTM. The encoding, retention, activity-silent, and retrieval time periods began at 270 ms, 2125 ms, 3883 ms, and 5055 ms after the memory set, respectively. For each TOI, 200 time series are constructed of variable length starting at next 200 time points of the time series at a temporal resolution of 256 Hz after the start of the TOI. Therefore, for each TOI, the LSTM has an average of 200*390 = 78,000 time series samples for analysis. This step is crucial due to the high number of weight parameters the LSTM contains as well as to prevent overfitting. All of the time series samples are down sampled further to a temporal resolution of 256/4 = 64 Hz. This temporal resolution is high enough to capture the temporal dynamics of EEG which is mainly characterized by activities less than 30 Hz. Time series lengths varied from length 1 to length 57. This means that the actual time scale of analysis ranged from 1/64 s to 57/64 s.

The LSTM can decode working memory load from neural signals in several ways: (1) Information from individual electrode level such as mean and variance, (2) information from the pattern based relationship between electrodes, (3) information from the joint temporal dynamics of multiple electrodes. We standardized the CSD data so that we can focus on pattern based information and sequential information. Information from the individual electrode level is not robust and subject to noise. The LSTM can decode neural activity on pattern based information and sequential information. We wanted to develop a method of assessing the contribution of sequential information to decoding memory load. Therefore, the LSTM decoding was executed with two difference scenarios, ordered and temporally shuffled. The procedure for the order scenario has been described already. In the shuffled scenario, all the time series samples are temporally shuffled randomly for each fold so that the input data has no temporal structure. However, the pattern based information between the electrodes is preserved. This forces the LSTM to rely on pattern-based spatial information rather than sequential temporal information to decode memory load. We investigated the deviation in the performance between the two scenarios to assess the contribution of sequential information to represent WM information.

## LSTM Weight Analysis

The weights from the input layer to the LSTM cells for each of the gates can be mapped topographically to determine the importance of an EEG electrode for the gate's role in decoding working memory load. For each EEG electrode and gate, the absolute value to the weights is averaged across LSTM cells and folds. This produces a 128 value topographical plot on the scalp for each of the gates. This process is completed for ordered and shuffled cases. The average absolute value of weights assesses the contribution of the EEG electrode for decoding working memory load because weights with a larger deviation from zero allow the EEG input node to have a larger dynamic range for influencing the LSTM network's classification. The topographical plots from the learned weights in the length 35 EEG time series chosen as higher lengths yielded marginally diminishing increases in decoding accuracy. The difference between the topographical plot for the ordered case and shuffled case is computed to determine which EEG electrodes have the most deviation in information contribution to decoding between the cases. A Z-score of this plot is computed to capture the topographical variation in the deviation between ordered and shuffled cases for the EEG node contribution to decoding.

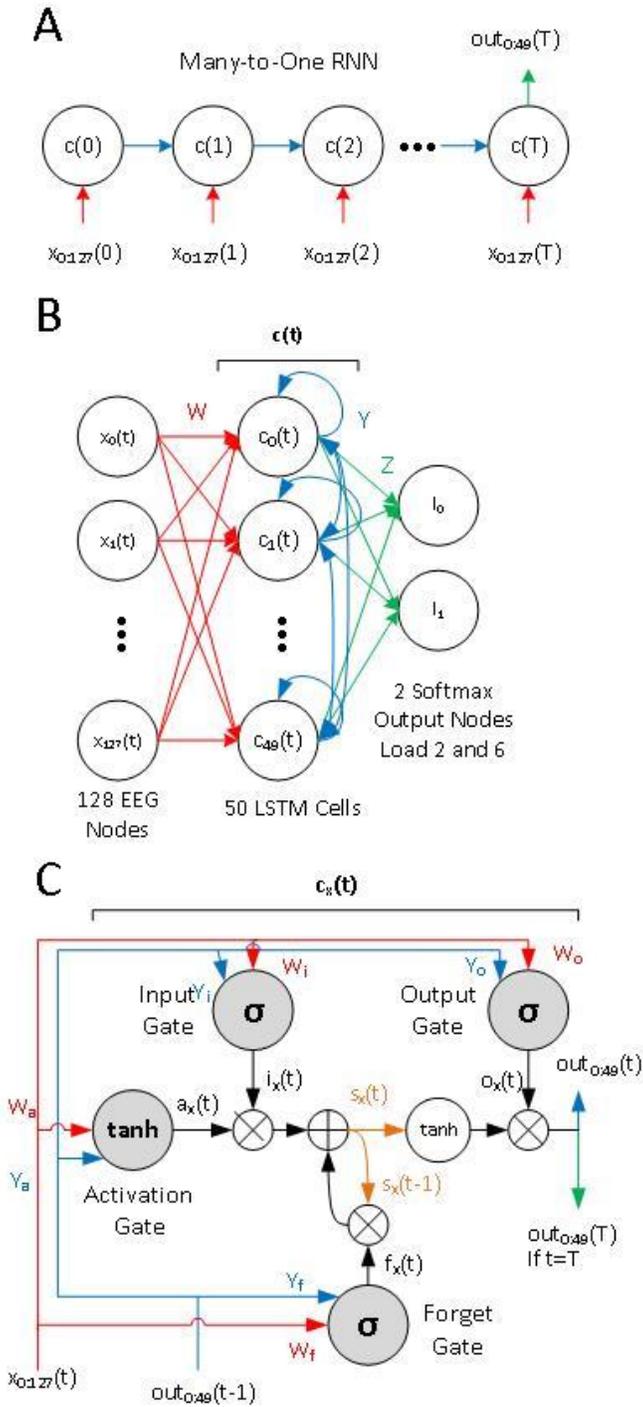

$t \in [0, T)$ (1)

$\sigma(u) = \frac{1}{1+e^{-u}}$ (2)

$a(t) = \tanh(W_a * x(t) + Y_a * out(t-1) + b_a)$ (3)
[50x1]     [50x128] [128x1] [50x50] [50x1]    [50x1]

$i(t) = \sigma(W_i * x(t) + Y_i * out(t-1) + b_i)$ (4)
[50x1]   [50x128] [128x1] [50x50] [50x1]    [50x1]

$f(t) = \sigma(W_f * x(t) + Y_f * out(t-1) + b_f)$ (5)
[50x1]   [50x128] [128x1] [50x50] [50x1]    [50x1]

$o(t) = \sigma(W_o * x(t) + Y_o * out(t-1) + b_o)$ (6)
[50x1]   [50x128] [128x1] [50x50] [50x1]    [50x1]

$s(t) = a(t) \cdot i(t) + f(t) \cdot s(t-1)$ (7)
[50x1]  [50x1]  [50x1]  [50x1]   [50x1]

$out(t) = \tanh(s(t)) \cdot o(t)$ (8)
[50x1]       [50x1]      [50x1]

$l = softmax(out(T)' * Z)$ (9)
[1x2]         [1x50]    [50x2]

$l(u) = softmax(out(T)' * Z(u)) = \frac{e^{out(T)'*Z(u)}}{\sum_{u=0}^{1} out(T)'*Z(u)}$ for $u \in 0,1$ (10)

Figure 2. LSTM networks. (A) A many-to-one RNN is used where multiple time points contribute to memory load classification. (B) Architecture of the LSTM networks. (C) LSTM cell. Orange array represents the internal state of the cell which recurrently updates as time points are fed into the LSTM. (D) Equations corresponding to LSTM.

# Results

## SVM-based Decoding

As shown in Figure 3, after presentation of the memory set (also referred to as cue), SVM decoding accuracy increases to a local maximum around ~550 ms post cue. The first TOI, corresponding to the encoding of the memory set, was based on this time. Decoding accuracy decreases towards the end of the encoding period. After the offset of the cue (2 second post cue), the decoding accuracy starts to increase again, reaching a local maximum centered at 400ms post cue offset. The second TOI, corresponding to the retention of working memory, is chosen around this time. Decoding accuracy decreases as the retention period progresses until reaching chance level at about 1800 ms post cue offset. This time period, called the activity-silent memory period in which we were no longer able to decode working memory load from EEG data, is the basis for choosing the third TOI. After the probe onset, decoding accuracy starts to increase yet again, reaching a local maximum centered at 650 ms post probe onset. The fourth TOI, corresponding the retrieval of working memory, is chosen according to this time.

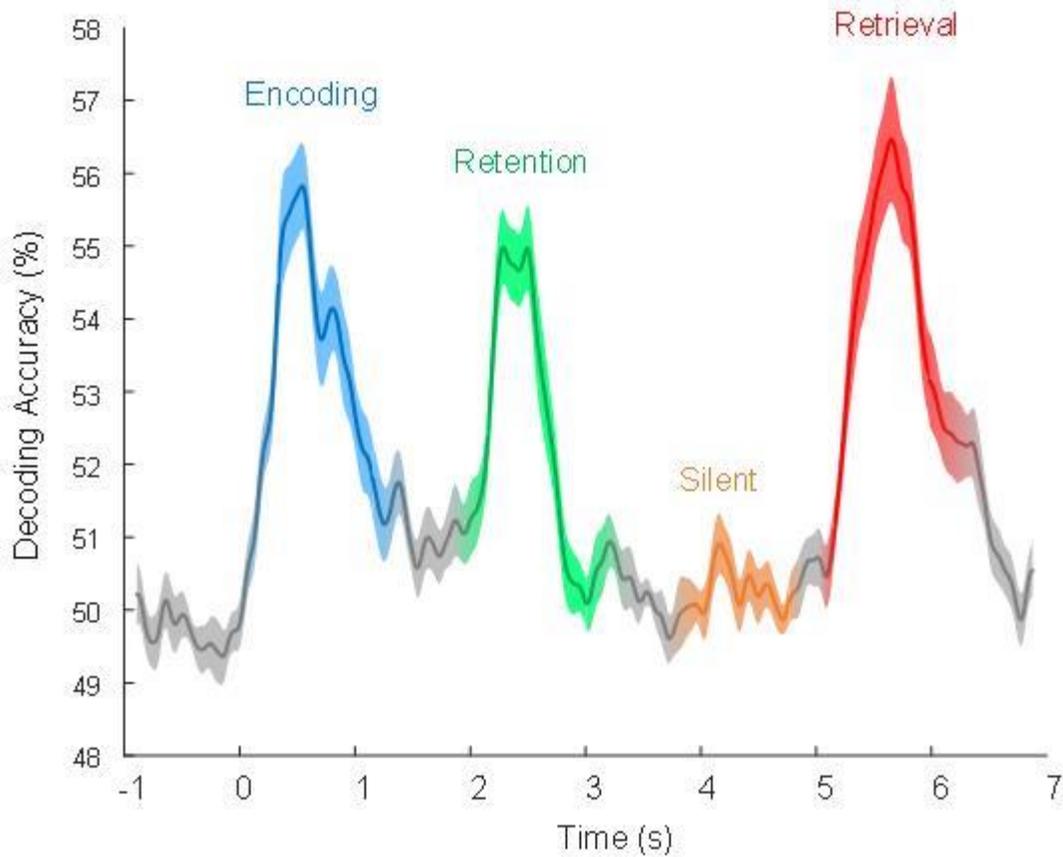

Figure 3. Time course of decoding accuracy for load 2 vs load 6 conditions with SVM during the working memory task. Decoding accuracy is used to assign time periods of interest for the four memory stages in which time series are constructed for LSTM decoding. Shaded region shows the standard error across subjects.

## LSTM-based Decoding

As shown in Figure 4A, for all four TOIs, LSTM decoding accuracy increases with length of the input time series. The plateaued decoding accuracy is the highest during retention followed by encoding, retrieval and activity-silent period. The decoding accuracy of the ordered time series is higher than the shuffled time series, and this difference increases with increase in the length of the time series for all four TOIs (Figure 4B). However, there is no significant difference between ordered and shuffled accuracies for the activity-silent period. The decoding accuracy during encoding and retrieval show the largest deviation between ordered and shuffled scenarios.

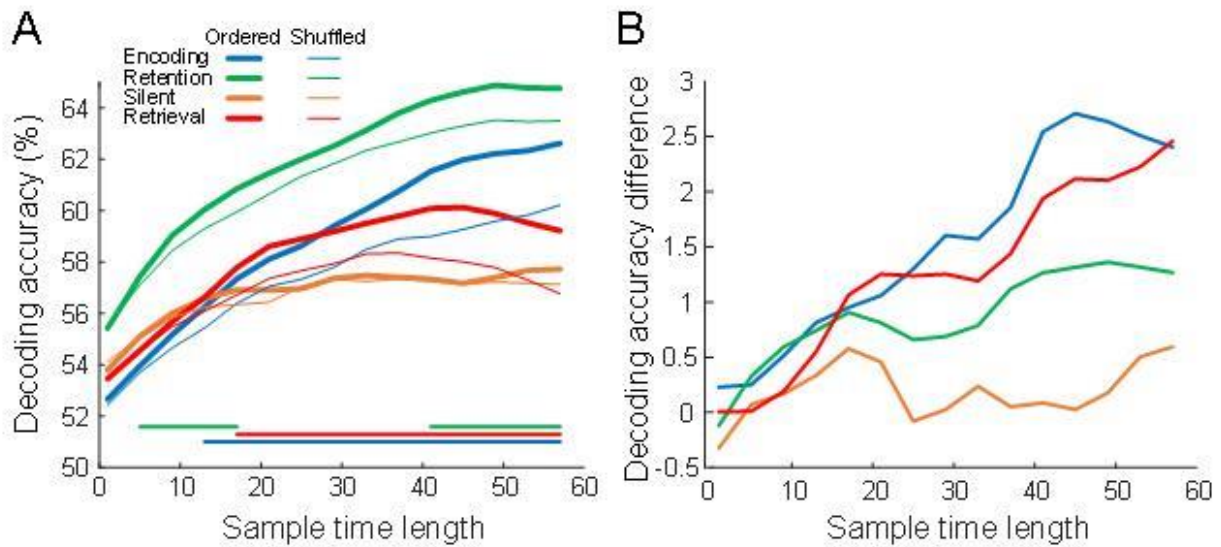

Figure 4. LSTM decoding. (A) LSTM decoding accuracy for all memory stages across different input time series lengths. The decoding accuracy of the ordered data (thicker line) is compared to the temporally shuffled data (thinner line). The horizontal lines at the bottom represent time lengths associated with decoding accuracies that are significantly different via a two-sample t-test ($p < 0.05$) in the ordered and shuffled scenario. (B) The difference between the decoding accuracy of the ordered and shuffled data across variable time lengths.

Although removing the temporal structure of the data in the shuffled case yielded statistically significant lower decoding accuracy, it did not drastically change the decoding accuracy as one might expect.

## LSTM Weight Map

We analyzed the weight maps of the LSTM after training to investigate how the network learns to differentiate working memory load from the temporally ordered data versus the temporal shuffled data. This may offer additional insight into the contribution of sequential information to decoding WM load as well as into the contribution of different brain regions. It is evident that the network in the shuffled case self organizes differently to compensate for the lack of temporal sequential information by learning more spatial pattern-based information. This can explain why the decoding accuracy is not dramatically different between the ordered and the shuffled cases. However, the weight maps show a dramatic difference in how they self-organize.

According to the weight map results in Figure 5, the absolute value of the weights is generally higher in frontal, temporal, and some occipital and parietal areas. In Figure 5A, the absolute values of the weights are especially lower in the central regions for the Input gate. In the ordered scenario, it is clear that the encoding weight map generally has higher average absolute values than the weight maps from other stages of the task.

The topographical plots of the absolute value of the forget gate weights are generally higher in the ordered case than the unordered case. However, the topographical plot of the absolute values of the input gate has higher absolute values in the shuffled case than the ordered case. Therefore, the forget gate weights learn more information in the ordered case when sequential information is available while the input gate weights learn more information when only spatial pattern based information is available. This occurs because the input gate weight map self organizes to learn spatial pattern information while the forget weight map self organizes to learn more temporal information. In the shuffled scenario when there is no temporal dynamics to learn, the forget gate weight map learns less information; however, the input gate weight map compensates by learning more spatial pattern information. In Figure 2D (Equation 7), the input gate and forget gate compete with each other to input new information into the cell or keep current temporal information, respectively.

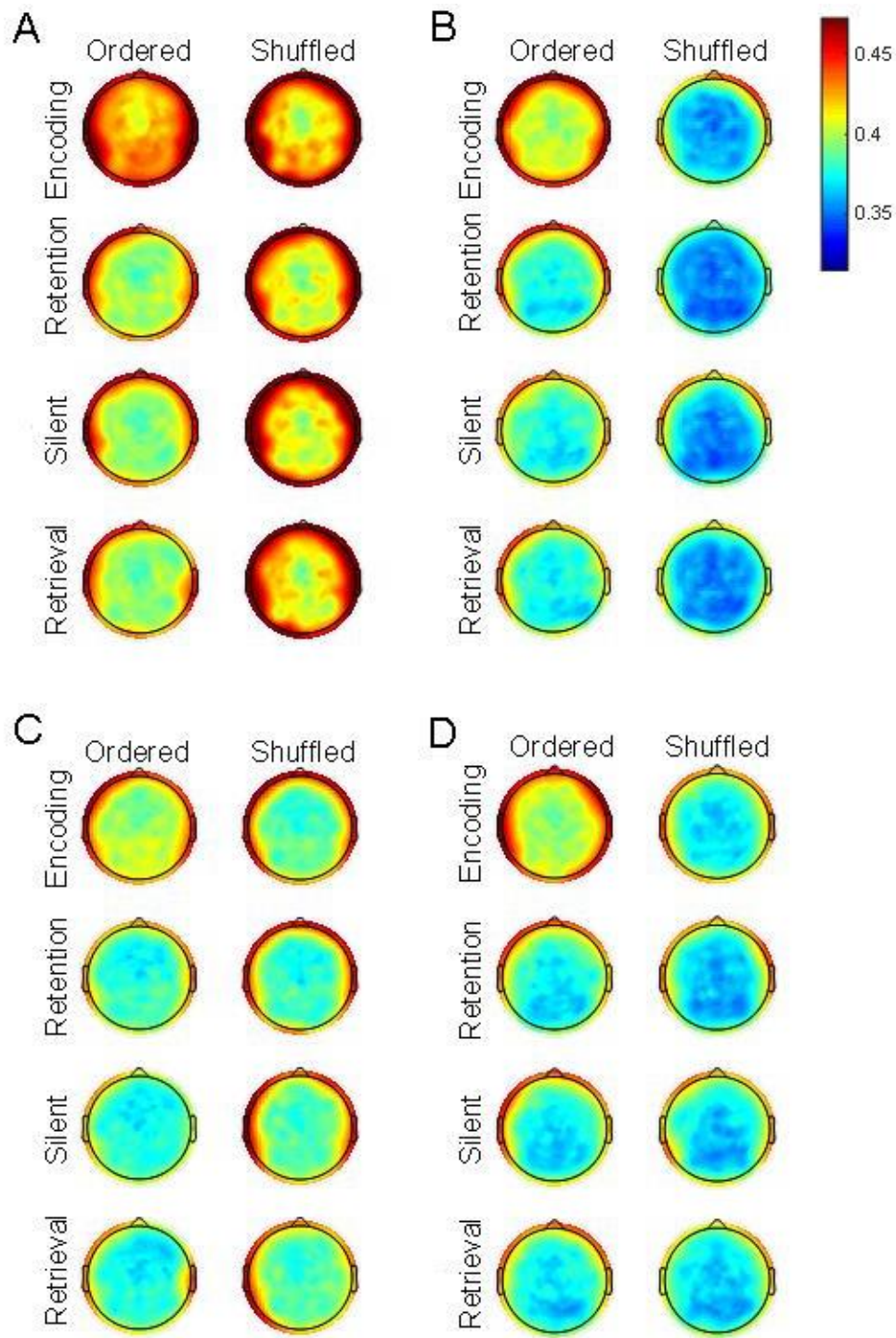

Figure 5. The average absolute value for the final weights from the input layer to the LSTM layer are topographically mapped for ordered and temporally shuffled input data for all memory stages. The weights correspond to the following LSTM Gates: (A) Input (B) Forget (C) Activation (D) Output.

The absolute value of the input gate weights is higher than the forget gate weights, regardless of whether the data is ordered or temporally shuffled. This is true because it is easier for the LSTM to learn spatial pattern information in general than temporal sequential information. The activation gate has similar behavior as the input gate for this compensation argument where it has higher absolute weights in the shuffled case than ordered case. However, the difference in the activation of the two gates, shown in Figure 2D (Equations 3 and 4), accounts for the difference in how they self-organize.

It is important to note that these weights discussed are only weights from the input nodes to the LSTM cells (Figure 2, Red). The recurrent weights among the LSTM cells (Figure 2, Blue), all self-organize to learn more in the ordered case than the shuffled case according to a 2-sampled t-test between average absolute value of ordered and shuffled weights ($p < 0.05$). However, these weights are not analyzed further in this study, as they cannot be mapped topographically. As it makes the most sense to analyze the weights from the input nodes to the LSTM cell for the input gate and forget gate due to the discussion above, the output gate weights from the input nodes to the LSTM cell are less interpretable mathematically and are excluded from further discussion.

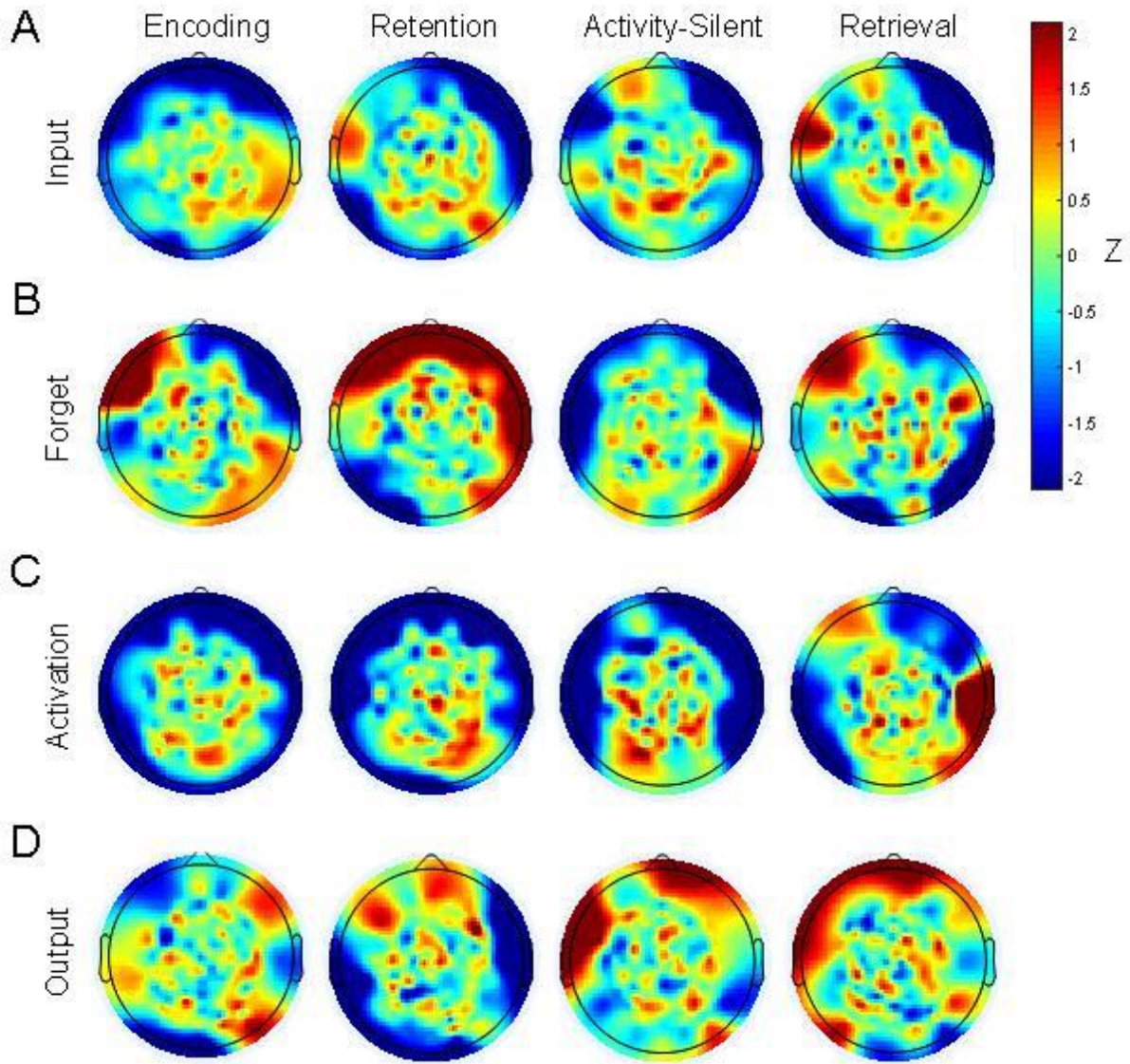

Figure 6. For each memory stage, the topographical plot of weights for the temporally shuffled scenario is subtracted from the topographical plot of weights for the ordered scenario. A z-score of the topographical plot is computed.

Figure 6 shows the deviation of the topographical plot between the ordered case and shuffled case. From the explanation above, only the Input gate (Figure 6A) and Forget gate (Figure 6B) will be discussed. A 2-sample t-test was completed between the ordered and shuffled forget weights for the following predefined ROIs: left frontal, right frontal, left temporal, right temporal, left parietal, right parietal, left occipital, and right occipital. All the ROIs show significantly higher absolute value of weights in the ordered case than shuffled case ($p < 0.05$). Therefore, forget weights self organize to learn more information in the ordered case than shuffled case for all brain areas. However, Figurew 6B can reveal brain regions that have a relatively increased deviation between ordered and shuffled weight maps.

According to Figure 6B, encoding and retrieval periods appear to have the most deviation between ordered and shuffled forget weight maps in the left frontal region. The retention period appears to have the most deviation in forget weight maps in the bilateral frontal and right temporal interestingly. The activity-silent period shows no significant increase in relative deviation in frontal regions. These deviations for the various memory stages are interpreted as indicating that increased sequential information contributes to decoding memory load. According to Figure 6A, the average absolute values of the weight maps have a relatively larger decrease in value between the ordered than shuffled scenario in the brain regions with a lower z-score. Therefore, compensatory spatial pattern information may be generally localized in frontal and occipital regions for the WM stages.

## Discussion

We recorded high-density EEG from healthy human volunteers performing a verbal WM task with different levels of memory load. Our purpose is to examine whether WM load information is encoded in EEG temporal dynamics. Using SVM to define four time periods of interest, which corresponded to different stages of the working memory process, we then applied LSTM-RNN to find that the decoding accuracy increases with increasing time series length for all WM processing stages. After randomly shuffling the time index of the original ordered time series, the decoding accuracy decreased when compared to the originally ordered time series, but remained an increasing function of the length of time series.

Current decoding analysis of EEG data is done in such a way that at each time point a separate decoding model is fit to the data. Whether there is information in the temporal patterns of EEG data remains largely unexplored. Our results suggest that the memory-load specific information is encoded in ordered temporal patterns of EEG data and LSTM networks can decode such patterns. Moreover, even for the time-index shuffled time series, the LSTM has more information to decode load when more instances of temporally static patterns, therefore more features, are available. Interestingly, Figure 4B shows that the deviation between the ordered and the shuffled decoding accuracy increases with the increase in the length of time series data. This suggests that the sequential relationship between patterns of EEG activity becomes increasingly more informative in decoding memory load as longer samples of sustained temporal activity are provided to the LSTM.

The decoding accuracy shows the largest deviation between ordered and shuffled scenarios during encoding and retrieval period, suggesting that sequential dynamics plays a more important role during stimulus-evoked energetic states of encoding and retrieval than during less energetic states such as retention. However, retention in general has higher decoding accuracy than encoding and retrieval. This may suggest that there is a larger deviation between load conditions in the sustained pattern based information of the EEG during the retention period even though sequential information is not as informative in the other memory processes.

In verbal WM, a key concept is the phonological loop, consisting of two components: a phonological store for temporary retention of verbal information and a subvocal rehearsal system that actively refreshes the decaying verbal information in the phonological store while having the ability to encode and retrieve verbal information (Baddeley, 2003). The subvocal rehearsal system and phonological store create a closed loop system that is required to maintains and manipulates information sequentially (Burgess and Hitch, 1999). The larger deviation in decoding accuracy between ordered and shuffled scenarios for the stimulus-evoked encoding and retrieval periods may be accounted for by the subvocal rehearsal system deploying more sequential serial processing. The retention period has less deviation in decoding accuracy between the ordered and shuffled cases. Later in retention, an activity-silent period was observed in which SVM decoding is at chance level. However, with increasing in the length of time series, LSTM is able to decode at above chance level, but there is no difference between the decoding accuracy for ordered and shuffled scenarios. As during the activity-silent period, WM information maintenance relies primarily in synaptic weights rather than sustained neural firing, sequential processing may play a lesser role (Mongillo et al., 2008; Stokes, 2015).

To assess the contributions of different electrodes to LSTM decoding, we plotted the weight maps as topographical plots. The forget gate weight maps generally have higher absolute value in the ordered scenario than in the shuffled scenario (Figure 5B). However, the input gate weight maps have higher absolute value in the shuffled scenario (Figure 5A). This may reflect that the input gate weight map learns a compensatory increase in pattern-based information when the temporal structure of the input samples is destroyed while the spatial structure is retained. Equation 7 (Figure 2D) supports that the input gate and forget gate have a tradeoff relationship for self-organizing to learn pattern-based information (input gate) or sequential information (forget gate) exclusively for input node to LSTM cell weights. Analyzing the deviation in the forget weight maps for the ordered and shuffled data may reveal brain regions where sequential information is more crucial for decoding WM load.

While scalp EEG is not a strong method for pinpointing the neuroanatomical basis of electrophysiological phenomena, the findings in Figure 6 showing that sequential information during encoding and retrieval primarily come from the left frontal cortex broadly agrees with the notion that the left frontal cortex, specifically the Broca's area, is important in the manipulation and maintenance of verbal information as part of the subvocal rehearsal system. The bilateral frontal and right temporal cortex appear to be an important neural substrate of sequential information in retention. Interestingly, the contribution of sequential information for decoding WM load in the frontal lobe is bilateral during retention; however, it is lateralized to the left during manipulation of WM load. The activity-silent period has minimal sequential information contribution to decoding as the information is not stored actively in sequential firing of neural populations but rather in synaptic weight mechanisms (Mongillo et al., 2008; Stokes, 2015). The parietal cortex contains attention control mechanisms that are required during all stages of the WM process. Some of the weight map topographies appear to be consistent with the involvement of these brain regions such as more sequential information coming from the left parietal cortex in encoding and retention.

This study, to our knowledge, is the first study applying a LSTM-RNN approach to investigate the contribution of sequential temporal information in human EEG to memory load decoding. Contrasting original order time series against randomly shuffled time series and mapping topographically the weights of different gates onto electrodes are additional innovative aspects of our approach.

## Conflict of Interest Statement

The authors declare that the research was conducted in the absence of any commercial or financial relationships that could be construed as a potential conflict of interest.

## Acknowledgments

This work was supported by the National Institute of Health grants MH112206.